# The structure of N184K amyloidogenic variant of gelsolin highlights the role of the H-bond network for protein stability and aggregation properties

Matteo de Rosa[1,2]*, Alberto Barbiroli[3], Francesco Bonì[1,2], Emanuele Scalone[1,2], Davide Mattioni[1,2], Maria A. Vanoni[2], Marco Patrone[4], Michela Bollati[1,2], Eloise Mastrangelo[1,2], Toni Giorgino[1,2] and Mario Milani[1,2]

[1]Istituto di Biofisica, Consiglio Nazionale delle Ricerche, via Celoria 26, 20133 Milano, Italy. [2]Dipartimento di Bioscienze, Università degli Studi di Milano, via Celoria 26, 20133 Milano, Italy. [3]Dipartimento di Scienze per gli Alimenti, la Nutrizione e l'Ambiente, Università degli Studi di Milano, Milano, Italy. [4]Biocrystallography Unit, Division of Immunology, Transplantation and Infectious Diseases, IRCCS San Raffaele Scientific Institute, via Olgettina 58, 20132 Milano, Italy.

*to whom correspondence should be addressed: matteo.derosa@cnr.it

## Abstract

Mutations in the gelsolin protein are responsible for a rare conformational disease known as AGel amyloidosis. Four of these mutations are hosted by the second domain of the protein (G2): D187N/Y, G167R and N184K. The impact of the latter has been so far evaluated only by studies on the isolated G2. Here we report the characterization of full-length gelsolin carrying the N184K mutation and compare the findings with those obtained on the wild type and the other variants. The crystallographic structure of the N184K variant in the $Ca^{2+}$-free conformation shows remarkable similarities with the wild type protein. Only minimal local rearrangements can be observed and the mutant is as efficient as the wild type in severing filamentous actin. However, thermal stability of the pathological variant is compromised in the $Ca^{2+}$-free conditions. These data suggest that the N to K substitution causes a local disruption of the H-bond network in the core of the G2 domain. Such a subtle rearrangement of the connections does not lead to significant conformational changes but severely affects the stability of the protein.

## **Keywords**

## **Acknowledgments**

This research was supported by a Research Grant of the Amyloidosis Foundation (Michigan, United States) awarded to MdR. The diffraction experiments were performed on beamline ID23-1, ID29 and Massif-3 at the European Synchrotron Radiation Facility (ESRF, France). We are thankful for the provided beamtime and assistance during the measurements.

# 1. Introduction

Gelsolin (GSN) is the prototype of a superfamily of $Ca^{2+}$-dependent proteins which regulate actin oligomerization (Nag et al. 2013) and possesses six homologous domains (named G1 to G6). These modules share a gelsolin-like fold, characterized by a 5- or 6-strand β-sheet sandwiched between two α helices, and host at least one $Ca^{2+}$ binding site. Alternative initiation of transcription and splicing produce two GSN variants: a cytoplasmic and an exported isoform of the protein. The mature form of the latter is 23 residues longer than that retained in the cell, and is used as a reference for the numbering of the residues throughout the manuscript. In the absence of $Ca^{2+}$, GSN adopts a compact and almost globular conformation (Burtnick et al. 1997; Nag et al. 2009). Domains G1 to G5 are wrapped around the central G6 and a C-terminal latch locks the protein in such conformation by binding to the G2. Actin binding interfaces are buried in the $Ca^{2+}$-free configuration which is therefore usually referred to as the *inactive* or *closed* GSN state. $Ca^{2+}$ affinity for the individual domains varies between 0.2 μM and 600 μM (Pope et al. 1995; Zapun et al. 2000; Chen et al. 2001; Khaitlina et al. 2004); in this concentration range, the domains unwind and GSN progressively adopts more open configurations, eventually leading to the fully *active* form (Kiselar et al. 2003; Ashish et al. 2007). The active/open conformation of GSN is a flexible structure, with the domains mostly connected by loose linkers (Ashish et al. 2007). For this reason, many structural and functional studies relied on the use of the isolated domains (Isaacson et al. 1999; Kazmirski et al. 2000, 2002; Ratnaswamy et al. 2001; Huff et al. 2003; Bonì et al. 2016, 2018; Giorgino et al. 2019). In addition to $Ca^{2+}$, other intracellular secondary messengers (Badmalia et al. 2017; Szatmári et al. 2018; Patel et al. 2018); as well as environmental factors, such as temperature and pH, (Garg et al. 2011; Badmalia et al. 2017) modulate GSN localization, conformation and/or activity.

Several mutations in the gelsolin protein have been described as responsible for AGel amyloidosis, a rare genetic disease first described in the '70s (Meretoja 1969). Among them, the D187N, D187Y, G167R and N184K substitutions are within the G2 domain (Fig. 1, numbering of the mature plasma protein) (Meretoja 1969; de la Chapelle et al. 1992; Sethi et al. 2013; Efebera et al. 2014). The D187N, D187Y substitutions impair $Ca^{2+}$ binding to G2, destabilize the G2 fold and increase its conformational flexibility (Fig. 1) (Kazmirski et al. 2000, 2002; Ratnaswamy et al. 2001; Chen et al. 2001; Huff et al. 2003; Bonì et al. 2016; Giorgino et al. 2019). In the Golgi, high concentrations of $Ca^{2+}$ activate GSN and the mutation-induced G2 destabilization promotes its aberrant cleavage by the furin protease (Chen et al. 2001). The larger proteolytic product, the C68 fragment, is exported in the extracellular space, where it is further processed by matrix-metalloproteases (MMPs) (Page et al. 2005). The 5 kDa and 8 kDa fragments produced by processing of D187N/Y variants are the toxic species which deposit as amyloid fibers in different districts of the body (Solomon et al. 2009; Efebera et al. 2014). This furin-dependent amyloidogenic pathway has been recently challenged, as it was shown that even undigested D187N GSN could form amyloid-like structures under acidic conditions (Srivastava et al. 2018). The N184K mutant of GSN, which is responsible for a renal AGel amyloidosis form, has been recently described and only partially characterized (Efebera et al. 2014; Bonì et al. 2016). The reasons behind the tissue-specificity of the GSN variants is not yet fully understood. G2 carrying the N184K substitution is still able to bind $Ca^{2+}$, and the geometry of the binding site matches that of the wild type (WT) protein (Bonì et al. 2016). Destabilization of the N184K G2 variant seems to be caused by the remodeling of the H-bond network in the core of the domain (Fig. 1). The different mechanism of mutation-induced destabilization of G2, however, causes susceptibility to furin proteolysis comparable to that induced by D187N/Y substitutions (Bonì et al. 2016), suggesting that the furin/MMPs proteolytic cascade is also responsible for the aggregation of the N184K variant.

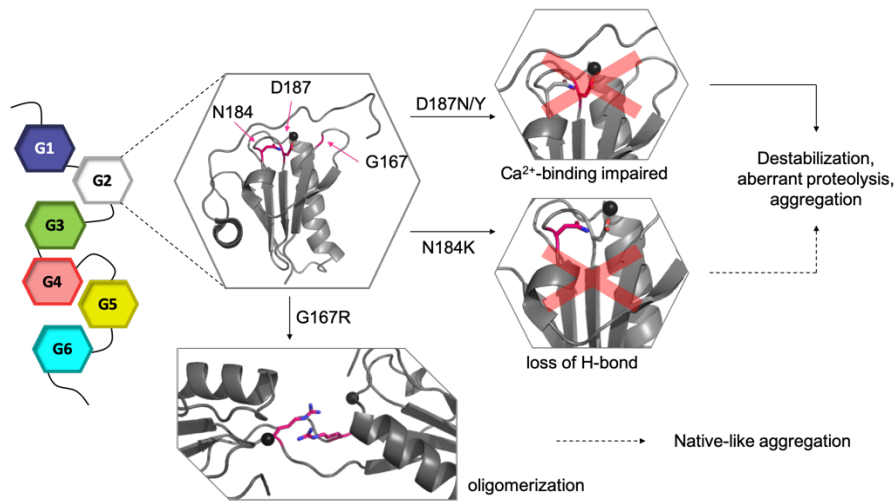

**Figure 1: Schematic representation of GSN architecture, localization and impact of some of the amyloidogenic mutations.** Amyloidogenic mutations hosted by G2 are mapped onto the crystallographic structure of the wild type form of the G2 domain (PDB ID: 6QW3) (Bollati et al. 2019). Dashed arrows indicate processes yet to be fully described.

## 2. Materials and methods

### 2.1 Protein production

WT full length human GSN and the three pathological variants (D187N, N184K and G167R) were expressed as heterologous proteins in *E. coli* cells and purified following the protocols reported in (Bonì et al. 2016, 2018; Giorgino et al. 2019).

### 2.2 Crystallization, structure solution and analysis

Crystallization trials were performed using an Oryx-4 nanodispenser robot (Douglas Instrument). Vapor-diffusion experiments were carried out at 20°C in sitting-drop set up. Full length GSN (in 20 mM HEPES, 100 mM NaCl, 1 mM EGTA, 1 mM EDTA, pH 7.4) was concentrated to 10 mg/ml (120 μM). Drops of 0.4 μl were prepared with different protein/precipitant ratios using several commercial screen solutions. The best diffracting crystals of D187N, N184K and G167R FL variants appeared under conditions similar to those previously reported (Nag et al. 2009; Zorgati et al. 2019). Crystals were further optimized in 1.3-1.5 M ammonium sulfate, 100 mM Tris-HCl, 20% glycerol, pH 8.0-8.5.

X-ray diffraction data of FL D187N, N184K and G167R were collected (at 100 °K) at beamline ID23-1, ID29 and Massif-3 (ESRF, Grenoble), respectively. Data were processed using XDS (Kabsch 2010) or DIALS (Clabbers et al. 2018) and scaled with AIMLESS (Evans and Murshudov 2013). Structures were solved by molecular replacement with PHASER (McCoy et al. 2007) using the WT gelsolin crystal structure (PDB ID 3FFN (Nag et al. 2009)) as a search model. Phenix refine (Afonine et al. 2018) was used for the refinement of the structures while manual model building was performed with Coot (Emsley et al. 2010). The structures of full length $Ca^{2+}$-free GSN variant D187N, N184K and G167R were deposited with PDB ID 6QBF, 6Q9R and 6Q9Z, respectively.

Analysis of the structures was performed with PyMOL (Schrödinger; DeLano, 2002), which was also used to prepare the figures, and LigPlot+ (Laskowski and Swindells 2011). Global RMSD and *per-residue* analysis were performed with ProSMART (Nicholls et al. 2014). B-factors were normalized (Bz-score) and analyzed as reported in (de Rosa et al. 2015).

**2.3 Thermal denaturation monitored by circular dichroism spectroscopy**

Thermal stability of GSN variants was evaluated as previously reported (Bonì et al. 2016, 2018; Giorgino et al. 2019). Briefly, proteins were diluted to 0.2 mg/ml in 20 mM HEPES, pH 7.4, 100 mM NaCl and either 1 mM EDTA or 1 mM $CaCl_2$. Loss of protein secondary structure was monitored at 218 nm during a 20 to 95 °C temperature ramp (1 °C/min). When appropriate, Tm values were calculated as the maximum of the first-derivative of the sigmoidal traces. N184K variant was assayed also at 0.1, 0.5 and 1 mg/ml to assess if aggregation kinetics affects apparent thermal stability

**2.4 Fluorimetric actin severing assay**

Pyrene-labeled rabbit skeletal muscle globular actin (G-actin; Cytoskeleton, Inc., (Denver, CO, USA)) was used to evaluate the severing activity of the pathological variants of gelsolin. Preparation of the solutions, actin manipulation and conversion of G-actin to filamentous actin (F-actin) were performed as reported elsewhere (Vanoni et al. 2013; Vitali et al. 2016). Measurements were performed at 20 °C with a Cary Eclipse fluorimeter (Agilent Technologies, USA), with the following settings: excitation and emission wavelength/slit of 365/5 and 407/5 nm, respectively; averaging time 0.1 s. A 3 ml cuvette was used, hosting a stirring bar for continuous agitation of the reaction mixture. Four hundred μl of a 4 μM F-actin solution were incubated in the cuvette until stabilization of the fluorescence signal, then 2 μl of 50 μM GSN were added (0.25 μM final concentration). Once the signal was stable again, the severing reaction was started by adding 2 μl of a 1 M $CaCl_2$ solution (final free $Ca^{2+}$ concentration > 1 mM). Data were normalized assuming that the fluorescence measured before addition of the protein corresponds to F-actin and that the fluorescence intensity of G-actin is equal to 0 arbitrary units.

# 3. Results

## 3.1 N184K mutation causes minimal rearrangements of the protein in the $Ca^{2+}$-free conformation

Crystals of the full-length N184K mutant and, for comparison, those of the D187N and G167R variants, were obtained under conditions known to stabilize WT-GSN in the closed/inactive conformation, which include, beside the absence of $Ca^{2+}$ ion, careful control of temperature and pH (here 20 °C and pH 8.0-8.5). Quality of the diffraction data for the N184K variant is comparable to that of previously published structures (Nag et al. 2009; Zorgati et al. 2019). The resolution of this dataset (2.7 Å, Table 1), although modest, allows an unbiased comparison among all variants. In all crystals, two GSN molecules are present in the asymmetric unit. However, as the quality of the electron density is superior for chain A, this molecule is used throughout the paper for the analyses. Most of the residues belonging to the six homologous domains (G1-G6) are unambiguously traced while the flexible linkers, loosely connecting the modules, are often missing in the models. In particular, the following chain A residues were not modelled: 155-156, 259-263 and 373-377 in the D187N structure; 153-156 and 259-261 in the N184K structure; and 156-157, 259-263 and 618-621 in the G167R structure.

| **Dataset** | D187N | N184K | G167R |
|---|---|---|---|
| PDB id | 6QBF | 6Q9R | 6Q9Z |
| **Data collection** | | | |
| Space group and cell dimensions | | P 4 $2_1$ 2 169, 169, 151; 90, 90, 90 | |
| a, b, c; α, β, γ (Å;°) | | | |
| Number of unique reflections | 28,491 | 58,899 | 22,724 |
| Resolution range (Å) | 75.88-3.49 (3.69-3.49) | 75.26-2.73 (2.80-2.73) | 48.47-3.80 (4.10-3.80) |
| I/σ(I) | 4.4 (2.0) | 8.3 (2.1) | 8.1 (2.2) |
| CC 1/2 | 0.948 (0.265) | 0.996 (0.353) | 0.993 (0.751) |
| Completeness (%) | 99.5 (96.6) | 99.9 (100.0) | 99.9 (100.0) |
| Multiplicity | 10.3 (8.1) | 9.1 (9.4) | 11.3 (11.9) |
| **Refinement** | | | |
| Resolution range (Å) | 75.88-3.49 | 75.26-2.73 | 48.47-3.80 |
| Number of reflections | 28,300 | 58,852 | 22,686 |
| Total number of atoms | 11,208 | 11,503 | 11,169 |
| Rwork/Rfree* | 0.218/0.276 | 0.213/0.260 | 0.216/0.264 |
| RMSD Bonds/angles (Å/°) | 0.002/0.551 | 0.004/0.708 | 0.003/0.582 |
| Ramachandran outliers (%) | 0.4 | 0.8 | 0.9 |
| B factors (Å²) § | 74.0 | 61.0 | 109.0 |

*Table 1: Data collection and refinement statistics of full-length proteins.* **Values in parentheses refer to the highest resolution shell.** *$R_{work} = \Sigma_{hkl}||Fo| - |Fc|| / \Sigma_{hkl}|Fo|$ for all data, except 5-10%, which were used for $R_{free}$ calculation. § Average temperature factors over the whole structure.

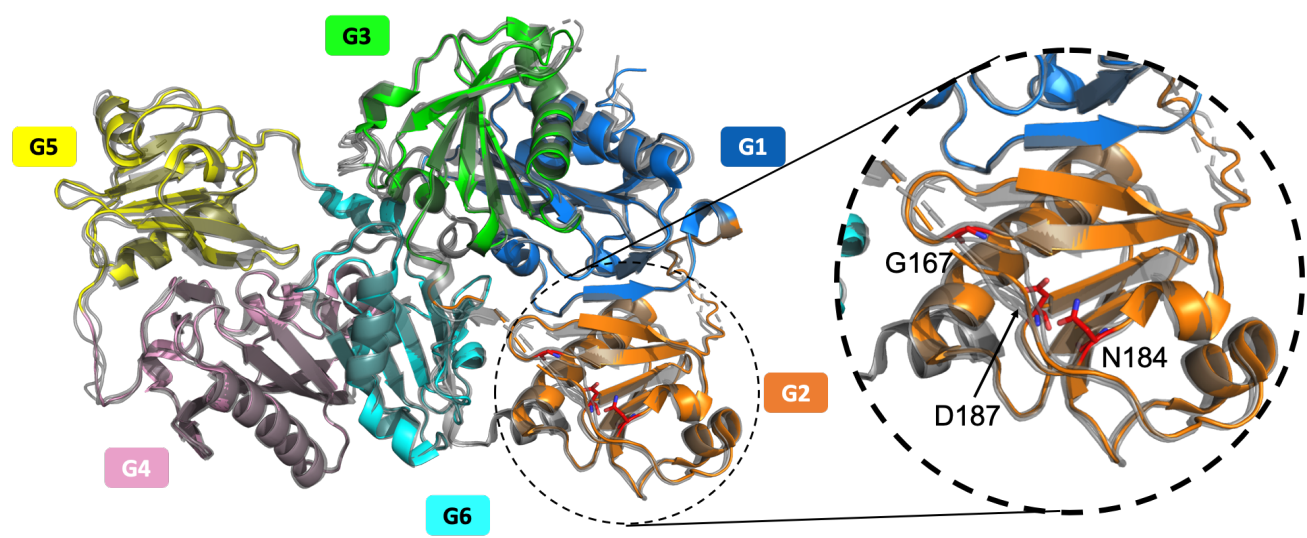

**Figure 2: GSN overall structure and domain organization.** *Cartoon representation of the full-length WT GSN in the $Ca^{2+}$-free conformation. Each domain is labelled and identified by a different colour. The structure of the D187N, N184K and G167R variants, all coloured in grey, are superimposed on that of the WT protein. Unresolved stretches are indicated by dashed lines while the residues substituted in the three pathological variants, are visualized as red sticks.*

| RMSD (Å) FL/G2 vs. | WT chain B | D187N chain A | N184K chain A | G167R chain A |
|---|---|---|---|---|
| **WT** chain A | 0.74/0.37 | 0.92/0.51 | 0.81/0.55 | 0.89/0.63 |
| **D187N** chain A | | | 0.72/0.48 | 0.74/0.71 |
| **N184K** chain A | | | | 0.83/0.81 |

*Table 2: Impact of the mutations on the overall structure of GSN. Global RMSD values (Å) for the full length structures, calculated from the alignment of either 720 Cα atoms of the full-length proteins (FL=G1 to G6) or 101 Cα atoms of the G2 domain extracted from the models of the full-length protein forms.*

A superimposition of the models of the four variants is shown in Fig. 2. Visual inspection of the superimposed structures does not reveal any major difference in the protein architecture. This observation is confirmed by the values of the root-mean squared distances (RMSD) of Cα atoms, calculated pairwise either over the whole chain or limited to the G2 residues of the GSN variants (0.72-0.92 and 0.48-0.81Å, respectively), which are comparable to those between the two asymmetric WT molecules (0.74 and 0.37 Å) (Table 2).

Amyloidogenic mutations often lead to subtle structural differences which are still sufficient to (partially) misfold the protein and trigger its aggregation. To assess such differences, *per-residue* RMSD was calculated using the WT protein (PDB ID 3FFN (Nag et al. 2009)) as the reference. This analysis revealed several RMSD peaks largely shared by all the variants (data not shown). To understand whether these differences are ascribed to conformational changes or to the increased dynamics of the corresponding stretch of the protein, we analyzed the correlation between B-factors and RMSD. Most of the RMSD peaks correlate with high B-factors and fall in the linker regions between adjacent gelsolin-like domains. These linkers are poorly resolved in the structures, and often the main chain was not modeled. Therefore, these RMSD and B-factor differences appeared to be due to *mutation-independent* intrinsic flexibility of the interdomain loops.

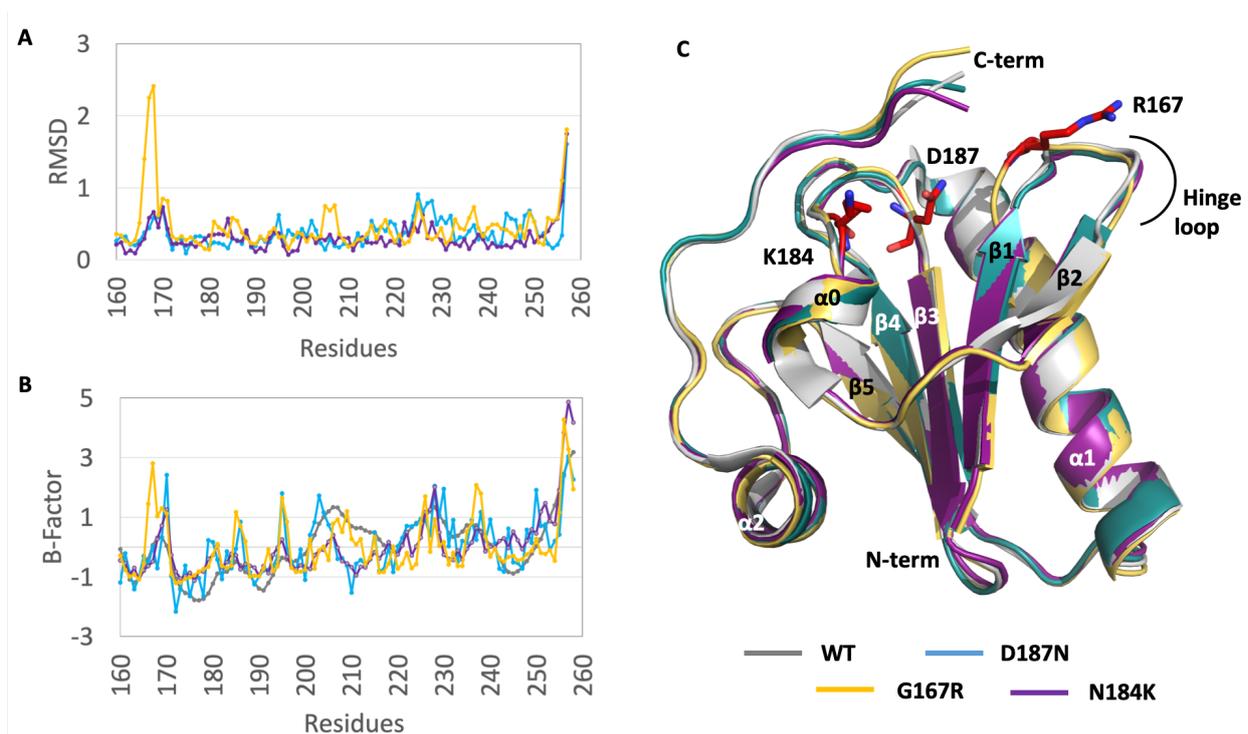

*Figure 3: Structure and dynamics of G2 portions in the $Ca^{2+}$-free conformation.* *RMSD (A) and differential (mutant-WT) normalized B factor (B) for the G2 domain (residue 160-258) from the $Ca^{2+}$-free FL structures. (C) superimposition of the same G2 portion of the full-length structures (WT, D187N, N184K and G167R), the mutated residues are displayed as red sticks and labeled.*

Local differences between the full-length WT protein and the D187N or G167R variants have been recently discussed elsewhere (Zorgati et al. 2019). Thus, hereafter they will be only briefly summarized. In the absence of $Ca^{2+}$, the D187N mutation has no impact on the structure of the hosting G2 domain. The D-to-N substitution is isosteric and the loss of charge, which is essential for ion binding, is well tolerated in a cavity otherwise crowded with negatively charged residues. In the G2 of the G167R variant, a sharp RMSD peak is observed corresponding to the first half of the hinge loop (residues 166-168; Fig. 3A). These high values do not correlate with higher normalized B factors, suggesting a *bona fide* conformational change rather than increased flexibility (Fig. 3B). Indeed, the hinge loop presents a kink in correspondence of the mutation, which is required to release the torsional strain induced by the G-to-R replacement.

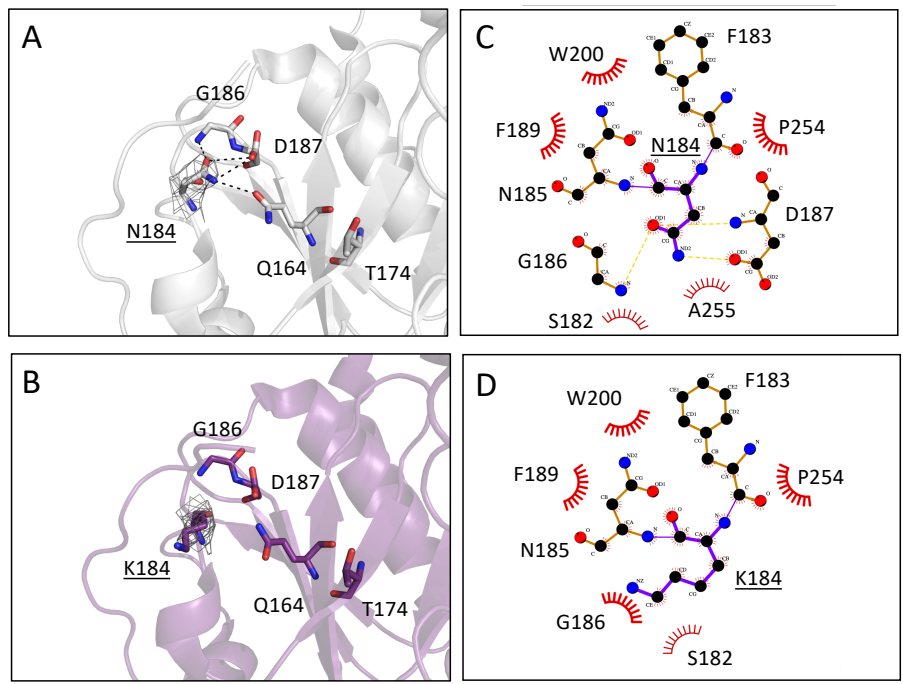

*Figure 4: Local rearrangements in the structure of FL $Ca^{2+}$-free GSN induced by the N184K mutation*. *(A,B) Close-up view of the area of the region harboring residue 184 in the WT (A) and N184K (B) protein. The substituted residue and the residues either H-bound (as computed by PyMol, Schrödinger; DeLano, 2002) or observed in alternative conformations in the relative $Ca^{2+}$-bound structure (Bonì et al. 2016) are represented as sticks and labeled. Electron density*

*(contoured at 1.5 σ) for the residue at position 184 is also displayed. (C, D) Interactions of residue 184 in the WT (C) and N184K (D) GSN variants analyzed by LigPlot+ (Laskowski and Swindells 2011). H-bonds are displayed as dashed yellow lines and the corresponding residues as sticks, spoked arcs represent residues making non bonded contacts.*

In the crystallographic structure of the N184K variant, the longer and charged K184 side-chain is extruded from the core of the G2 domain and points toward the solvent (Fig. 4A,B). Analyses of the interactions of the 184 residue in the WT and N184K variant was performed with PyMOL and LigPlot+ (Laskowski and Swindells 2011). Both programs show that in the WT protein, N184 makes H-bonds with G186 and D187 (Fig. 4A,C) and that both interactions are lost in the N184K mutant (Fig. 4B,D). Additionally, PyMOL also detects a weaker (3.4 Å-distance) H-bond between N184 and Q164, as well absent in the mutant. A similar rearrangement of the H-bond network was also observed in the crystallographic structure of the $Ca^{2+}$-bound G2 carrying the same amino acyl substitution (Bonì et al. 2016). Thanks to the higher resolution of this structure, it was shown that the loss of connection leads to a destabilization of the G2 C-terminal stretch and the hinge loop (residues 167-172). In the context of the full-length protein, it could further cause the weakening of the intimate connection between G1 and G2 domains, and, as a consequence, an increased overall instability of the protein (see below).

**3.2 N184K mutation significantly decreases GSN stability but does not impair its physiological activity**

Impact of the amyloidogenic mutations on the thermodynamic stability of GSN has been extensively investigated. However, most of these studies were performed on the isolated G2 domain and only recently some data on the full-length protein were reported (Srivastava et al. 2018; Zorgati et al. 2019). In the isolated domain, G167R and N184K mutations significantly decreases the stability of both the $Ca^{2+}$-free and $Ca^{2+}$-bound forms. On the contrary, D187N/Y substitutions show Tm values similar to that of the WT protein in the absence of the ion, because $Ca^{2+}$ binding is compromised (Isaacson et al. 1999; Ratnaswamy et al. 2001; Kazmirski et al. 2002; Huff et al. 2003; Giorgino et al. 2019). The thermal stability of the N184K variant of full-length GSN has not yet been studied yet, while that of the D187N, D187Y and N184K GSN variants was evaluated in the presence of a fluorogenic probe (Zorgati et al. 2019). Here we report circular dichroism (CD)-monitored (at 218 nm) unfolding experiments of WT, D187N, G167R and N184K, either in the presence of saturating $CaCl_2$ concentration (1 mM) or in the absence of the ion (1 mM EDTA). The $Ca^{2+}$-free form of GSN in its compact arrangement behaves as a globular protein and the melting curve shows high cooperativity and a signle transition between native and denatured states (Fig. 5, top left panel). As previously reported (Bonì et al. 2018; Zorgati et al. 2019), under this condition the G167R variant shows a significantly lower Tm of 51.5 °C as compared to the wild-type protein (Tm, 56.6 °C). A similar value measured for the N184K variant (Tm=51.9 °C, apparently not affected by protein concentration) was somehow unexpected based on the structural analysis, where only minor and local rearrangements were observed. Thermal stability of the D187N variant is also compromised (Tm=54.5 °C), but to a much lesser extent.

In the presence of a high concentration of $Ca^{2+}$ ions (> 0.5 mM), GSN adopts the fully open/active form, where many interdomain contacts are lost. Indeed, in the curves recorded under these conditions a loss of cooperativity was observed and at least two unfolding events were identified (Fig. 5, bottom left). Overall, all the mutants showed nearly overlapping curves. The first stage is still characterized by a steep decrease of the CD signal with a Tm of about 56 °C, while the second event (Tm, 61 °C) is more gradual. Since the secondary structure content is well conserved, we expect the 6 homologous domains to equally contribute to the CD signal, making interpretation of these curves relatively difficult. Moreover, temperature is known to induce the activation of GSN similarly to $Ca^{2+}$. These findings might suggest that the mutations have no impact on GSN in the $Ca^{2+}$-bound form, but studies on the isolated domain clearly showed that all the substitutions impairs G2 stability (Bonì et al. 2016, 2018; Giorgino et al. 2019). Therefore, we propose that in the $Ca^{2+}$-free form the local destabilizing contribution of the mutations is amplified by the cooperativity of the process. On the contrary, in the denaturation of the open form, where all the domains are likely independent, the mutation-induced destabilization of the G2 is concealed by the presence of the other 5 domains.

AGel amyloidosis has been historically associated solely to a *gain-of-toxic function*. That is, the mutations somehow trigger protein aggregation and the (pre)aggregates are toxic for the organism, resulting in an autosomal dominant disease. Consistently with this hypothesis, the full length D187N variant was shown to retain its ability to bind and sever actin filaments (F-actin) (Nag et al. 2009). The full length D187N, D187Y and G167R variants were also tested in actin-polymerization and -cosedimentation assays by others (Zorgati et al. 2019). Indeed, they all showed an efficiency in binding and promoting actin polymerization similar to that of the WT protein. We here investigated the F-actin severing activity of the N184K variant in fluorometric assays using pyrene-labeled actin. This assay was also designed to test if the mutations could activate the protein in a $Ca^{2+}$-independent fashion. Therefore, the F- to G-actin transition was initially monitored by following the time-course and extent of fluorescence decrease in the absence of $Ca^{2+}$ ions (Fig. 5, right panel). None of the tested variants showed any activity in these conditions. Upon addition of 1 mM free $Ca^{2+}$ ions, a steep decrease of the fluorescence was observed for every mutant with comparable efficiency. Our data together with the previous ones suggest that not only the N184K variants but all the so far-characterized amyloidogenic mutations do not impair F-actin severing and capping.

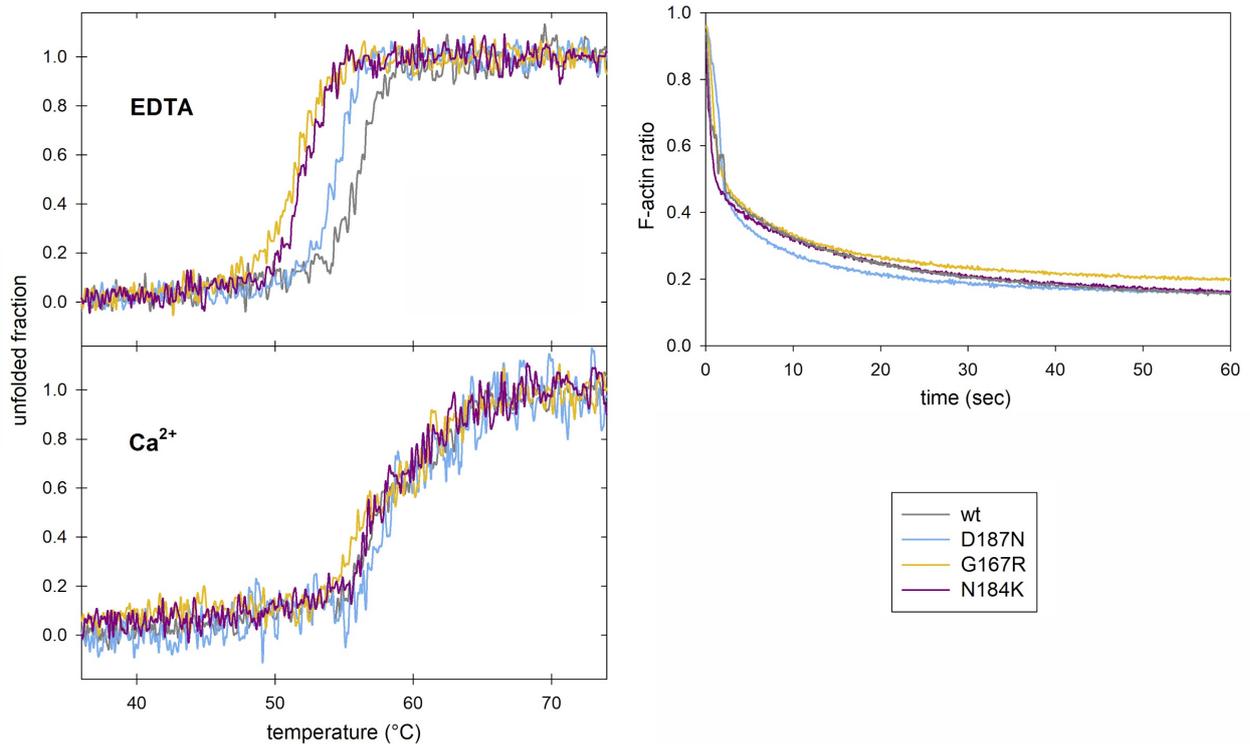

*Figure 5: Thermal stability and actin severing activity of WT and mutated GSN variants. Left) The thermal stability of WT, D187N, G167R and N184K GSN variants was evaluated by CD spectroscopy either in the presence of 1 mM EDTA or 1 mM $Ca^{2+}$. Right) The impact of the pathological mutations (WT, D187N, G167R and N184K) on the physiological activity of gelsolin was evaluated in severing assays, using pyrene-labelled F-actin.*

## 4. Conclusions

Several molecules and environmental factors potentially induce large conformational changes in gelsolin and modulate its dynamics. The mutual rearrangement of the domains is triggered by subtle local changes, which are not fully understood yet. Overall, the protein's dynamics is characterized by a striking balance between local and global effects that sometimes go in opposite directions. For example, $Ca^{2+}$ reduces the conformational flexibility of the single domains, but the full length $Ca^{2+}$-bound protein becomes very dynamics. Owing to technical difficulties, pathological GSN variants had been long studied mostly as isolated G2 domains and, only recently, efforts to validate the data on the full-length proteins were reported (Srivastava et al. 2018; Zorgati et al. 2019). In this context, we investigated structural and functional aspects of the GSN N184K variant. The newly described three-dimensional model of the variant in the $Ca^{2+}$-free conformation is in agreement with the previously reported atomic resolution structure of the N184K G2 domain (Bonì et al. 2016). The long and charged side-chain of lysine cannot be positioned in the cleft originally occupied by the asparagine. Thus, K184 is forced to reorient toward the solvent. Such rearrangement has no impact on the local geometry, but residue 184 loses its H-bonds with the neighboring Q164, G186 and D187 residues, in the core of G2. This is a severe loss of intradomain connection that, however, does not cause structural differences in the crystal structure. The broken connections likely increase the conformational flexibility of the protein, which becomes more vulnerable to thermal denaturation as observed by the significant decrease of the Tm values measured both on the mutated FL protein and, previously, on the isolated G2 (Bonì et al. 2016). These data indicate, again, an effect of the mutation on the dynamics of the protein rather than on its structure. This conclusion is also supported by the unaltered actin severing efficiency of the N184K variant. Since the G2 domain contributes to one of the actin binding interfaces, significant conformational changes would have likely caused differences in actin binding affinity or efficiency of severing.

AGel amyloidosis is yet an incurable disease and, owing to its rarity, relatively few resources have been devoted to find cures or strategies to ameliorate the symptoms. The data we presented here, along with those of other research groups, demonstrate that all of the pathogenic GSN variants characterized so far share structural and functional features. They should be taken into account and, possibly, exploited for the development of novel therapeutic strategies that aim to cure all forms of the disease.

# Conflict of interest

The authors declare that they have no conflict of interest.